\documentclass[letterpaper, 10 pt, journal, twoside]{ieeetran}  

\usepackage{amsmath,amssymb}
\usepackage{hhline}
\usepackage{bm}
\usepackage{color,soul}
\usepackage{graphicx}
\usepackage{caption}
\usepackage{subcaption}
\usepackage{longtable}	
\usepackage{multicol}
\usepackage{float}
\usepackage{multirow}
\usepackage{epsf,psfrag,amssymb,amsfonts,color,cite,fancybox}
\usepackage[mathscr]{eucal}
\IEEEoverridecommandlockouts

\usepackage{atbegshi}
\AtBeginDocument{\AtBeginShipoutNext{\AtBeginShipoutDiscard}}

\begin{document}

\title{Wide-Area Damping Control for Interarea Oscillations in Power Grids Based on PMU Measurements}

\author{Ilias Zenelis, Xiaozhe Wang, \textit{Member, IEEE}}

\thanks{This work is supported by Natural Sciences and Engineering Research Council (NSERC) Discovery Grant (NSERC RGPIN-2016-04570) and Fonds de Recherche du Qu\'{e}bec –- Nature et technologies (FRQ-NT NC-253053).

Ilias Zenelis and Xiaozhe Wang are with the Department of Electrical and Computer Engineering, McGill University, Montr\'{e}al, QC  H3A 0G4, Canada. email: ilias.zenelis@mail.mcgill.ca, xiaozhe.wang2@mcgill.ca}
\pagestyle{empty}
\maketitle
\thispagestyle{empty}

\begin{abstract}
In this paper, a phasor measurement unit (PMU)-based wide-area damping control method is proposed to damp the interarea oscillations that threaten the modern power system stability and security. Utilizing the synchronized PMU data, the proposed almost model-free approach can achieve an effective damping for the selected modes using a minimum number of synchronous generators.
Simulations are performed to show the validity of the proposed wide-area damping control scheme.
\end{abstract}

\section{INTRODUCTION}

\IEEEPARstart{L}{ow-frequency} interarea oscillations, involving two coherent generator groups swinging against each other at a frequency typically less than $2$ Hz, lead to a small-signal stability concern for the modern inter-connected power systems.
The undesirable existence of interarea oscillations due to weakly-tied transmission lines may limit the power transmission capability between different areas and damage the power grid elements, and therefore needs to be constantly monitored and controlled. Conventionally, power system stabilizers (PSSs) have been employed to damp the interarea oscillations. However, classical PSSs are incapable of damping the iterarea modes, the frequencies of which lie beyond their limited bandwidth.
Although a vast amount of techniques aiming to improve the traditional PSSs have been proposed,
including the multiple-input PSSs (e.g.,\mbox{\cite{Grondin98}}), multi-band PSSs (e.g.,\mbox{\cite{Lajoie98}}) and supervisory level PSSs (e.g.,\mbox{\cite{Ni02}}), PSS techniques may not effectively damp the interarea modes involving different areas and may strongly depend on the assumed network model\mbox{\cite{Bose08}}.

The implementation of a synchrophasor-based wide-area measurement system (WAMS) in  power grids greatly enhances the  observability of power system dynamics, 
providing a unique opportunity to observe, identify and damp the interarea oscillations. 
Multiple control methodologies have been developed for damping the interarea oscillations deploying a WAMS. A comprehensive discussion about the formulation of wide-area control problem in power systems was presented in\mbox{\cite{Chakrabortty13}}.
In \cite{Kamwa01}, a decentralized/hierarchical architecture for wide-area damping control using PMU remote feedback signals was discussed. In\mbox{\cite{Dorfler14}}, a sparsity-promoting optimal wide-area control was employed to damp the interarea oscillations in bulk power systems. References\mbox{\cite{Bose08}}\mbox{\cite{Raoufat16}} proposed the design of wide-area damping controllers (WADCs) that provide supplementary damping control to synchronous generators. The authors of \cite{Meng12} applied a networked control system model for wide-area closed-loop power systems. The authors of\mbox{\cite{Preece13}} introduced a power oscillation damping (POD) controller based on a WAMS using a modal linear
quadratic Gaussian (MLGC) methodology. However, these approaches require the detailed and accurate knowledge of the complete network model (both topology and parameter values), that is unavailable or corrupted in practice as a result of communication failures, bad data in state estimation etc. In addition, the impact of disturbances on the interarea oscillations cannot be well captured by these methods.

In this paper, we attempt to develop a wide-area damping control strategy for interarea modes utilizing PMU measurements, which does not rely on the assumed network model (the only required knowledge is the damping and inertia constants of generators which are not subject to constant changes). 
To the knowledge of authors, the proposed wide-area damping method for interarea oscillations seems to be the first method that is completely independent of the network model and its parameter values.
The main contributions of the paper are as below:
\begin{itemize}
\item A measurement-based (almost model-free) technique is applied to accurately estimate the system state matrix in ambient conditions, which is completely independent of the system network model and is computationally efficient.  
\item An effective wide-area damping control scheme for interarea modes is proposed using the participation factors of the estimated system state matrix, which can damp a target mode by a desired coefficient using the least possible number of generators while maintaining the other modes unaffected.
\item Numerical studies are conducted in the IEEE 39-bus 10-generator New England system to show that the proposed wide-area damping control method is fast, effective, and robust again measurement noise.
\end{itemize}


\section{THE STOCHASTIC POWER SYSTEM MODEL}\label{1}
In this paper, we investigate the power system dynamic operation in quasi steady-state, i.e., in ambient conditions. Since interarea modes are predominantly determined by the machine rotor angles and speeds, 
classical swing equations are used to model generator dynamics: 
\begin{equation}
\label{eq:class_gen}
\begin{gathered}
\dot{\delta}_{i} = \omega_{i}-\omega_{S} \\
M_{i}\dot{\omega}_{i} = P_{m_{i}}- P_{e_{i}} - D_{i}(\omega_{i}-\omega_{S}) \quad  i=1,...,n
\end{gathered}
\end{equation}
\normalsize
where $\delta_{i}$ is the generator rotor angle, $\omega_{i}$ is the rotor angular velocity, $\omega_{S}$ is the synchronous speed, $M_{i}$ is the inertia constant, $D_{i}$ is the damping coefficient, $P_{m_{i}}$ is the generator's mechanical power input from the prime mover, $P_{e_{i}}$ is the generator's electrical power output, and $n$ is the number of generators in the system.
$P_{e_{i}}$ is defined as
\begin{equation}
\label{eq:electric_power}
P_{e_{i}} = E_{i}\sum\limits_{j=1}^n E_{j}Y_{ij}\cos(\delta_{i}-\delta_{j}-\phi_{ij}) \quad i=1,...,n
\end{equation}
where $E_{i}$ is the constant voltage behind the transient reactance $X'_{d}$, and $Y_{ij}\angle{\phi_{ij}}$ is the $\textit{(i,j)}^{th}$ entry of the reduced admittance matrix containing generators' impedances. It should be pointed out that in \mbox{\eqref{eq:class_gen}} each generator represents the equivalent aggregation of thousands of actual generators.
%
\par
In the power system model of \eqref{eq:class_gen}-\eqref{eq:electric_power}, the loads are modeled as constant impedances. However, other types of loads such as ZIP loads can be incorporated in this formulation 
It is common and reasonable to assume that the load power varies stochastically following a Gaussian distribution\cite{Singh10}.
The load fluctuation manifests itself in the diagonal elements of the reduced admittance matrix as proposed in \cite{Bialek17},\cite{Zenelis2018}:

\begin{equation}
\label{eq:load_var}
Y(i,i) = Y_{ii}(1 + \sigma_{i}dW_{t,i})\angle{\phi_{ii}}
\end{equation}
\normalsize
where $W_{t}$ is a Wiener process and $\sigma_{i}$ is the standard deviation of the variation describing load fluctuations. Therefore, the power system equations become:
\begin{equation}
\label{eq:fluct_gen}
\begin{gathered}
\dot{\delta}_{i} = \omega_{i}-\omega_{S} \\
M_{i}\dot{\omega}_{i} = P_{m_{i}}- P_{e_{i}} -D_{i}(\omega_{i}-\omega_{S}) - E_{i}^{2}G_{ii}\sigma_{i}\xi_{i} 
\end{gathered}
\end{equation}
where $G_{ii} = Y_{ii}\cos(\phi_{ii})$, and $\xi_{i} = \frac{dW_{t,i}}{dt}, i=1,...,n$, are independent Gaussian random variables.
\par
Note that \eqref{eq:fluct_gen} represents a set of stochastic differential equations. To conduct the small-signal stability analysis,  we linearize \eqref{eq:fluct_gen} around the steady-state operating point as shown below:
\begin{equation}
\label{eq:matrix_form}
\begin{gathered}
\begin{bmatrix}
    \dot{\bm{\delta}}\\
    \dot{\bm{\omega}}
\end{bmatrix}
= \begin{bmatrix}
  0 & I_{n}\\
  -M^{-1}\frac{\partial \bm{P_{e}}}{\partial \bm{\delta}} & -M^{-1}D
\end{bmatrix}
\begin{bmatrix}
    \bm{\delta}\\
    \bm{\omega}
\end{bmatrix}
+\begin{bmatrix}
0\\
-M^{-1}E^2G\Sigma
\end{bmatrix}
\bm{\xi}
\end{gathered}
\end{equation}
where $\bm{\delta}=[\delta_{1},...,\delta_{n}]^T$, $\bm{\omega}=[\omega_{1}-\omega_{S},...,\omega_{n}-\omega_{S}]^T$, $M=\mbox{diag}([M_{1},...,M_{n}])$, $D=\mbox{diag}(D_1,...,D_n)$, $\bm{P_{e}}=[P_{e_{1}},...,P_{e_{n}}]^T$, $E=\mbox{diag}([E_{1},...,E_{n}])$, $G=\mbox{diag}([G_{11},...,G_{nn}])$, $\Sigma=\mbox{diag}([\sigma_{1},...,\sigma_{n}])$, and $\bm{\xi}= [\xi_{1},...,\xi_{n}]^T$.\\
Let $\bm{x}=[\bm{\delta},\bm{\omega}]^T$,
$A=\left[\begin{array}{cc}{{0}}&{I_n}\\-M^{-1}\frac{\partial{\bm{P_e}}}{\partial{\bm{\delta}}}&-M^{-1}D\end{array}\right]$, $B=[0,-M^{-1}E^2G\Sigma]^T$, then (\ref{eq:matrix_form}) takes the following compact form:
\begin{equation}
\dot{\bm{x}}=A\bm{x}+B\bm{\xi}
\end{equation}
In short, the stochastic power system dynamic model in ambient conditions can be represented as  
a vector Ornstein-Unlenbeck process that is Gaussian and Markovian. 
It will be discussed in Section \ref{method} that the dynamic system state matrix $A$ can be estimated from the statistical properties of the PMU measurements, based on which a measurement-based wide-area damping control scheme is developed.
\section{PMU-BASED WIDE-AREA DAMPING CONTROL}\label{method}\label{2}
\subsection{An (Almost) Model-Free Approach of Estimating $A$}

Assuming that the state matrix
$A$ is stable (satisfied in ambient conditions), the stationary covariance matrix $C_{xx}$ satisfies the following Lyapunov equation \cite{Gardiner09}:
\begin{equation}
\label{eq:lyapunov}
AC_{xx} + C_{xx}A^T = -BB^T
\end{equation}
%
where
$C_{xx} = \begin{bmatrix}
    C_{\delta\delta} & C_{\delta\omega}\\
    C_{\omega\delta} & C_{\omega\omega}
\end{bmatrix}$.
Equation (\ref{eq:lyapunov}) integrates the statistical properties of states that can be extracted from PMUs and the model knowledge, providing an ingenious way to estimate the model information from measurements.

Supposing that PMUs are installed at all the generator terminal buses (optimistic currently, yet not unreasonable in the near future), we can use the PMU measurements to calculate the values of rotor angle $\bm{\delta}$ and rotor speed $\bm{\omega}$ in ambient conditions as discussed in many previous works (e.g., \cite{Zhou11}). We, therefore, can further estimate the covariance matrix $C_{xx}$ of $\bm{\delta}$ and $\bm{\omega}$ (see Appendix). If the damping $D$ and inertia constants $M$ are known, it has been shown in \cite{Bialek17} that the dynamic state Jacobian matrix $\frac{\partial \bm{P_{e}}}{\partial \bm{\delta}}$ can be estimated by the following equation derived from \eqref{eq:lyapunov}:
\begin{equation}
\label{eq:dyn_jacobi}
(\frac{\partial \bm{P_{e}}}{\partial \bm{\delta}}) = MC_{\omega\omega}C_{\delta\delta}^{-1} - DC_{\omega\delta}C_{\delta\delta}^{-1}
\end{equation}
Importantly, we do not require any information about the network model (topology and parameter values) that is usually subject to inaccuracy due to, for instance, communication errors. Therefore, this method for estimating the dynamic state Jacobian matrix and the system state matrix is almost model-free (except the knowledge of $D$ and $M$). A brief overview of the detailed derivation of $\frac{\partial \bm{P_{e}}}{\partial \bm{\delta}}$ is presented in Appendix. Note that the conventional model-based method calculates the matrix $\frac{\partial \bm{P_{e}}}{\partial \bm{\delta}}$ by differentiating (\ref{eq:electric_power}) with respect to $\bm{\delta}$ that heavily depends on the network topology and parameter values embedded in the  admittance matrix $Y$.

Once the dynamic state Jacobian matrix $\frac{\partial \bm{P_{e}}}{\partial \bm{\delta}}$ is estimated, the system state matrix $A$ can be easily computed by:
\begin{equation}
\label{eq:state_matrix}
A = \begin{bmatrix}
  0 & I_{n}\\
  -M^{-1}\frac{\partial \bm{P_{e}}}{\partial \bm{\delta}} & -M^{-1}D
\end{bmatrix}
\end{equation}

\subsection{Modal Analysis and Linear Feedback Control}
The eigenvalues $\Lambda=\mbox{diag}([\lambda_1, ..., \lambda_{2n}])$ of $A$ appearing in complex conjugate pairs $\lambda_i = \eta_i\pm\omega_i, i =1,...,n$, the right eigenvectors $\Phi=[\phi_1,...,\phi_{2n}]$ and the left eigenvectors $\Psi=[\psi_1^T,...,\psi_{2n}^T]^T$ of $A$ can be readily extracted from the estimated matrix $A$.
Therefore, the mode frequencies $f_i=\frac{\omega_i}{2\pi}, i =1,...,n$ and the damping ratios $\zeta_i=\frac{-\eta_i}{\sqrt{{\eta_i}^2+{\omega_i}^2}}, i =1,...,n$ are straightforwardly obtained.
Moreover, the participation factor $P_i$ of $\lambda_{i}$ defined as:
\begin{equation}
P_i=[P_{1,i},...P_{2n,i}]^T=[\phi_{1,i}\psi_{i,1},...,\phi_{2n,i}\psi_{i,2n}]^T
\end{equation}
can be estimated from the right and left eigenvectors.

In addition, the matrix $\Lambda$ 
with the eigenvalues of $A$ as diagonal elements, can be written as:
\begin{equation}
\label{eq:diagonal_matrix}
\Lambda = {\Psi}A\Phi
\end{equation}
The left and right eigenvectors corresponding to $\lambda_{i}$ and $\lambda_{j}$ satisfy the following relation:
\begin{equation}
\label{eq:orthogonality}
\psi_j\phi_i =
\begin{cases}
  1, \quad \mbox{if } i = j \\
  0, \quad \mbox{if } i \neq j
\end{cases}
\end{equation}
where a vector normalization has been applied.

%

%
%

%
%
%

\subsection{The Proposed Wide-Area Damping Control Scheme}
In this paper, we intend to develop a wide-area damping control scheme using PMU measurements. Actually, we add a state feedback control loop to the original linear time-invariant open-loop system described by \eqref{eq:matrix_form} as shown below: 
\begin{equation}
\label{eq:feedback}
\begin{gathered}
\dot{\bm{x}} = A\bm{x} + B_{c}\bm{u}\\
\bm{u} = K\bm{x}\\
\end{gathered}
\end{equation}
where $\bm{x}=[\bm{\delta}, \bm{\omega}]^T$ is obtained from PMU measurements. The gain matrix $K$ is designed to damp the targeted interarea oscillation modes. The control center sends the input control signals $\bm{u}=K\bm{x}$ to the generators that participate in the WAMS-based central control as indicated by $B_c$.

The matrix $B_{c}$ is defined as:
$\small{
B_{c} = \begin{bmatrix}
    B_{c\delta} & 0\\
    0 & B_{c\omega}
\end{bmatrix}
}$,
%
where $B_{c\delta}$ and $B_{c\omega}$ refer to $\bm{\delta}$ and $\bm{\omega}$ respectively. Ideally, the remedial control scheme is applied to all $n$ generators 
and thus, $B_c = I$. However, it is rather impractical and expensive to apply a control measure to every synchronous machine. In this paper, the generators with the largest participation factors in regard to the mode of interest, are selected to conduct the damping control. Mathematically speaking, 
the diagonal entries of $B_c$ corresponding to the generators that no controls are carried out, are substituted by 0.


The closed-loop plant matrix $A_{cl}$ is given by:
\begin{equation}
\label{eq:clpm}
A_{cl} = A + \Delta A
\end{equation}
where $\Delta A = {B_c}K$, according to the state feedback loop defined in \eqref{eq:feedback}.
Representing $\Delta A$ in diagonal canonical form by applying the similarity transformation described by \eqref{eq:diagonal_matrix},
\begin{equation}
\label{eq:closed_diagonal}
\Delta\Lambda = \Psi\Delta A\Phi
\end{equation}
Hence, substituting $\Delta A = B_cK$ in the above relation,
\begin{equation}
\label{eq:closed_diagonal_final}
\Delta\Lambda = {\Psi}B_cK\Phi
\end{equation}

Inspired by the model-based damping technique introduced in \cite{Far09}, we propose a $2n\times2n$ damping matrix 
to damp the particular interarea oscillation mode $k$. In contrast to the model-based method in \cite{Far09}, the proposed wide-area damping control releases the dependence of the method on the accurate network model, topology, and parameter values, which are subject to frequent changes.
The subscript ($k$) is attached to the mathematical symbols thereafter to denote their reference to mode $k$. For instance, $K_{(k)}$ symbolizes the feedback matrix devoted to mode $k$. The two open-loop eigenvalues associated with mode $k$ are 
denoted as $\lambda_{k_1} = \eta_{k}+j\omega_{k}$ and $\lambda_{k_2} = \eta_{k}-j\omega_{k}$.
\subsubsection{Ideal Case}

As we have seen before, ideally $B_c = I$ if all generators receive the damping control signals. Assuming that we want to move the eigenvalues of mode $k$ (the conjugate pair $\lambda_{k_1}$ and $\lambda_{k_2}$) by a coefficient $\sigma_k<0$, we propose to use the following damping matrix:
\begin{equation}
\label{eq:gain_ideal}
K_{(k)} = \sigma_{k}[\phi_{k_1},\phi_{k_2}][\psi_{k_1}^T,\psi_{k_2}^T]^T
\end{equation}
Substituting \eqref{eq:gain_ideal} to \eqref{eq:closed_diagonal_final} with $B_c = I$, we have:
\begin{equation}
\label{eq:delta_lamda_ideal}
\Delta\Lambda_{(k)} = \sigma_{k}{\Psi}[\phi_{k_1},\phi_{k_2}][\psi_{k_1}^T,\psi_{k_2}^T]^T\Phi
\end{equation}
the $(i,j)^{th}$ entry of which is:
\begin{equation}
\label{eq:delta_lamda_ideal_3}
\Delta\Lambda_{ij(k)} = \sigma_{k}{\psi_i}\phi_{k_1}\psi_{k_1}\phi_j +\sigma_{k}{\psi_i}\phi_{k_2}\psi_{k_2}\phi_j
\end{equation}
Applying the orthogonality principle illustrated by \eqref{eq:orthogonality}, we have:
\begin{equation}
\label{eq:delta_lamda_ideal_4}
\Delta\Lambda_{ij(k)} =
\begin{cases}
  \sigma_k, \quad \mbox{if } (i,j)\in\{(k_1,k_1), (k_2, k_2)\}\\
   0, \quad \mbox{otherwise}
\end{cases}
\end{equation}
Therefore, $\lambda_{k_1} = \sigma_k+\eta_{k}+j\omega_{k}$ and $\lambda_{k_2} = \sigma_k+\eta_{k}-j\omega_{k}$, the 
eigenvalues of the closed-loop state matrix $A_{cl}$ corresponding to the targeted mode k, migrate to the left by a coefficient $\sigma_k$, leading to an improved damping. The rest of the eigenvalues remain unaffected under the proposed feedback damping control. 
\subsubsection{Practical Case}

In reality, $B_c$ may not be equal to $I$ as mentioned previously considering the cost of conducting control for all generators. If we still let the damping matrix $K$ to be:
\begin{equation}
\label{eq:gain_practical}
K_{(k)} = \sigma_{k}[\phi_{k_1},\phi_{k_2}][\psi_{k_1}^T,\psi_{k_2}^T]^T
\end{equation}
then
\begin{eqnarray}
\label{eq:delta_lamda_practical}
\Delta\Lambda_{(k)}&=& \sigma_{k}{\Psi}B_c[\phi_{k_1},\phi_{k_2}][\psi_{k_1}^T,\psi_{k_2}^T]^T\Phi\nonumber\\
& =& \sigma_{k}{\Psi}B_c\phi_{k_1}\psi_{k_1}\Phi +\sigma_{k}{\Psi}B_c\phi_{k_2}\psi_{k_2}\Phi\nonumber\\
&=&\sigma_{k}{\Psi}\hat{\phi}_{k_1}{\psi}_{k_1}{\Phi} +\sigma_{k}{\Psi}\hat{\phi}_{k_2}{\psi}_{k_2}{\Phi}
\end{eqnarray}
where
$
\hat{\phi}_{k_i} = B_c\phi_{k_i} \quad i = 1,2.\nonumber\\
$

As a result, $\hat{\phi}_{k_i}$ will have nonzero entries only if the corresponding generators  carry out the WAMS-based control.
For instance, presuming that only Generators 4-6 receive damping control signals, then we have
$\hat{\phi}_{k_i} =[\hat{\phi}^{\bm{\delta}}_{ k_i},\hat{\phi}^{\bm{\omega}}_{k_i}]= [0,0,0,{\phi}^{\delta_4}_{k_i},{\phi}^{\delta_5}_{k_i},{\phi}^{\delta_6}_{k_i},0,...,0,{\phi}^{\omega_4}_{k_i},{\phi}^{\omega_5}_{k_i},{\phi}^{\omega_6}_{k_i},0,...,0]^T$.
%

\begin{figure}[!ht]
\centering
\includegraphics[width=3.2in ,keepaspectratio=true,angle=0]{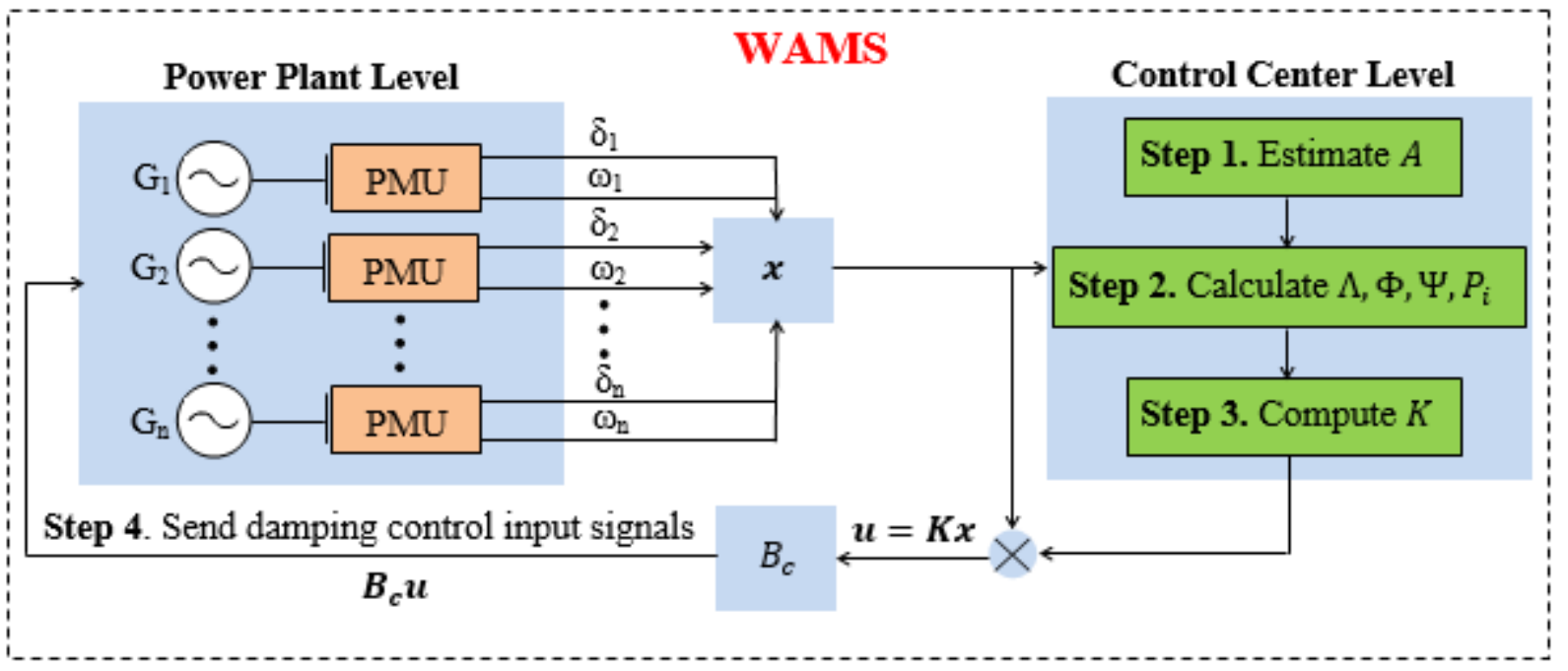}
\caption{Schematic diagram of the closed-loop system.}\label{closed_loop}
\end{figure}

By the eigenvalue perturbation theory \cite{Trefethen97}, the eigenvalues $[\hat{\lambda}_1,...,\hat{\lambda}_{2n}]$ of  $\Lambda+\Delta\Lambda_{(k)}$, and thus of $A+\Delta A_{(k)}$ satisfy:
\begin{eqnarray}
\hat{\lambda}_i&=&\lambda_i+e_i^T\sigma_k(\Psi\hat{\phi}_{k_1}{\psi}_{k_1}\Phi +\Psi\hat{\phi}_{k_2}{\psi}_{k_2}\Phi)e_i\nonumber\\
&=&
\begin{cases}
  \lambda_i+\sigma_k e_{i}^T\Psi\hat{\phi}_{i}, \quad \mbox{if } i\in\{k_1, k_2\}\\
  \lambda_i, \quad \mbox{otherwise}
\end{cases}
\end{eqnarray}
where $e_i$ denotes a unit vector that has $1$ in the $i^{th}$ position and 0 elsewhere. It is observed that although the damping effect to the targeted mode $k$ is slightly affected compared to the ideal case since  $e_{i}^T\Psi\hat{\phi}_{i}$ is typically different from 1, the other modes still remain unaffected. To ensure that an effective damping is acted to mode $k$ while minimizing the number of generators, we choose the generators with the largest participation factors in mode $k$ to carry out the control signals. The proposed WAMS-based damping control algorithm is presented below and is illustrated in Fig. \ref{closed_loop}. \vspace{4pt} \\
\textbf{Step 1.} Estimate the system state matrix $A$ using the PMU measurements by (\ref{eq:dyn_jacobi})-(\ref{eq:state_matrix}).\\
\textbf{Step 2.} Calculate the eigenvalues $\Lambda$, the right eigenvectors $\Phi$, the left eigenvectors $\Psi$ of the estimated $A$, and the participation factor $P_i$ for each mode $\lambda_{i}$. Select the interarea oscillation mode $k$ to damp.\\
\textbf{Step 3.} Compute the damping matrix $K$ by (\ref{eq:gain_ideal}) for the targeted mode $k$. \\
\textbf{Step 4.} Select the generators with the largest participation factors, find the corresponding $B_c$, and send the damping control input signals $\bm{u}={B_c}K\bm{x}$ to the selected generators.
\vspace{4pt}


In practice, the damping control signals are transmitted to the remote terminal units (RTUs) of the favored generators where they can serve either as ancillary control inputs to the generators' PSSs or as direct inputs to the generators' exciters. 

\section{NUMERICAL RESULTS}\label{3}

The IEEE 39-bus 10-generator New England system, is used to demonstrate the effect of the proposed control technique 
The topology of the system  
can be found in\mbox{\cite{Bialek17}}. For validation purposes, two case studies are presented. The first study intends to test the proposed control method on the classical generator models under which the method is developed. 
The second study is employed to demonstrate the validity of the suggested method in the real-world case where the generators are modelled as higher-order models and are controlled by exciters and PSSs. In addition, the PMU measurement noise is also considered.  All parameters for the two studies are available in: {https://github.com/zenili/Mode-Participation-Estimation-2017}. PSAT-2.1.9 \cite{Milano} is used for all simulations.


\subsection{Study I: Classical Generator Model}

The 10 generators are modelled as the classical model described by \eqref{eq:fluct_gen}. The angle of Generator 1 (G1) serves as the reference. The load fluctuations are characterized by a standard deviation $\sigma_i = 5$ in (\ref{eq:load_var}) for all generators. We assume that the sampling rate is 20 Hz, lying within the typical range of PMU sampling rate: 6-60 Hz\mbox{\cite{Kamwa01}}.
By executing the system state matrix estimation and modal analysis described in Section \ref{2}, all the eigenvalues can be estimated with a very good accuracy. Indeed, the estimation error is less than 2\% for frequencies and below 6\% for damping ratios for all modes. Specifically, Mode 7 that is characterized by the estimated values $f_7=1.662$ Hz (0.54\% estimation error) and $\zeta_7=1.03\%$ (5.56\% estimation error) is considered to be weakly damped as $\zeta_7<10\%$, which is a widely accepted criterion for satisfactory damping.



\begin{figure}[!ht]
\begin{subfigure}[t]{0.55\linewidth}
\includegraphics[width=1.8in ,keepaspectratio=true,angle=0]{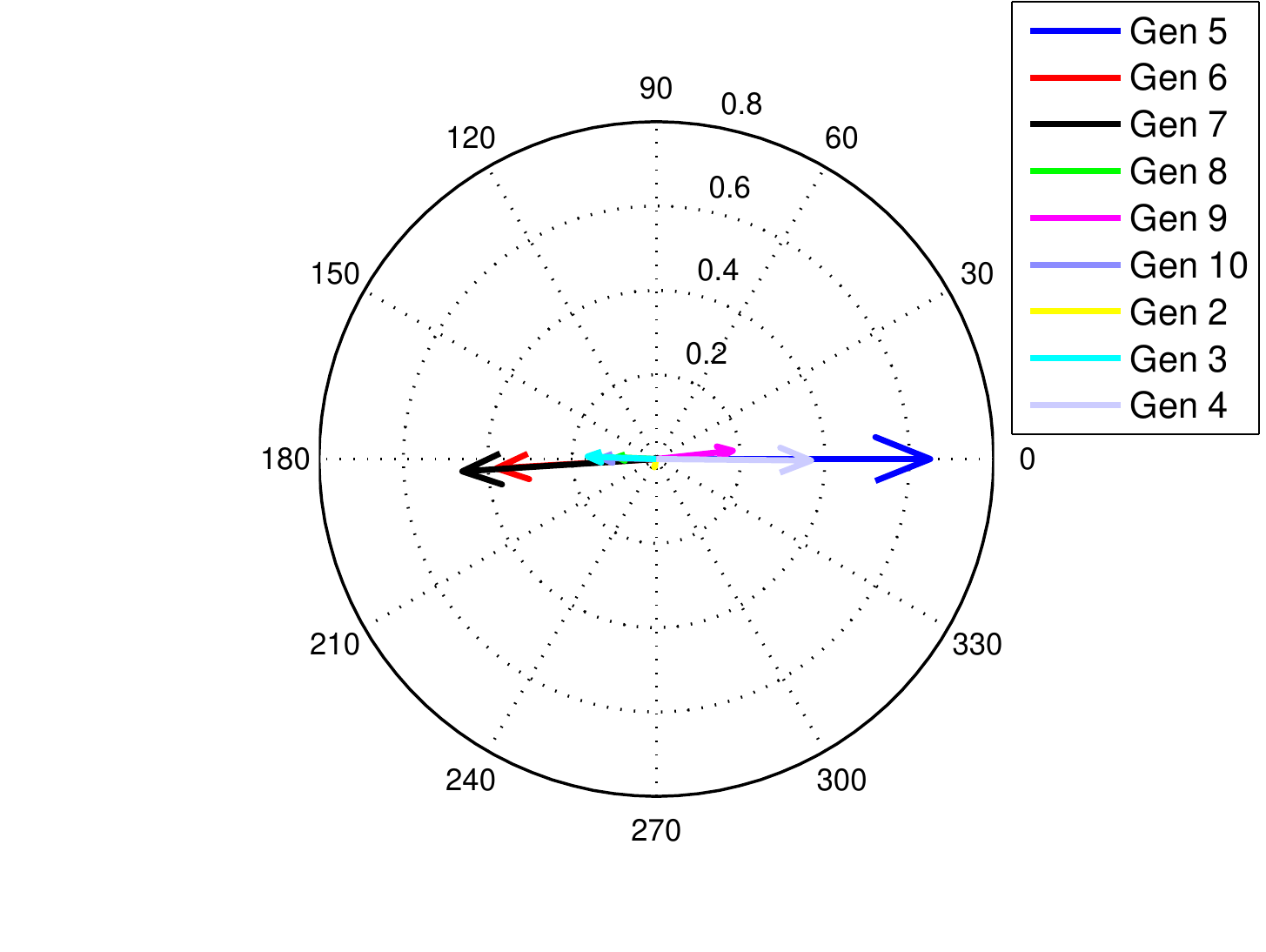}
\caption{Mode shapes for Mode 7.}\label{mode shape mode 7}
\end{subfigure}%
\begin{subfigure}[t]{0.45\linewidth}
\includegraphics[width=1.7in ,keepaspectratio=true,angle=0]{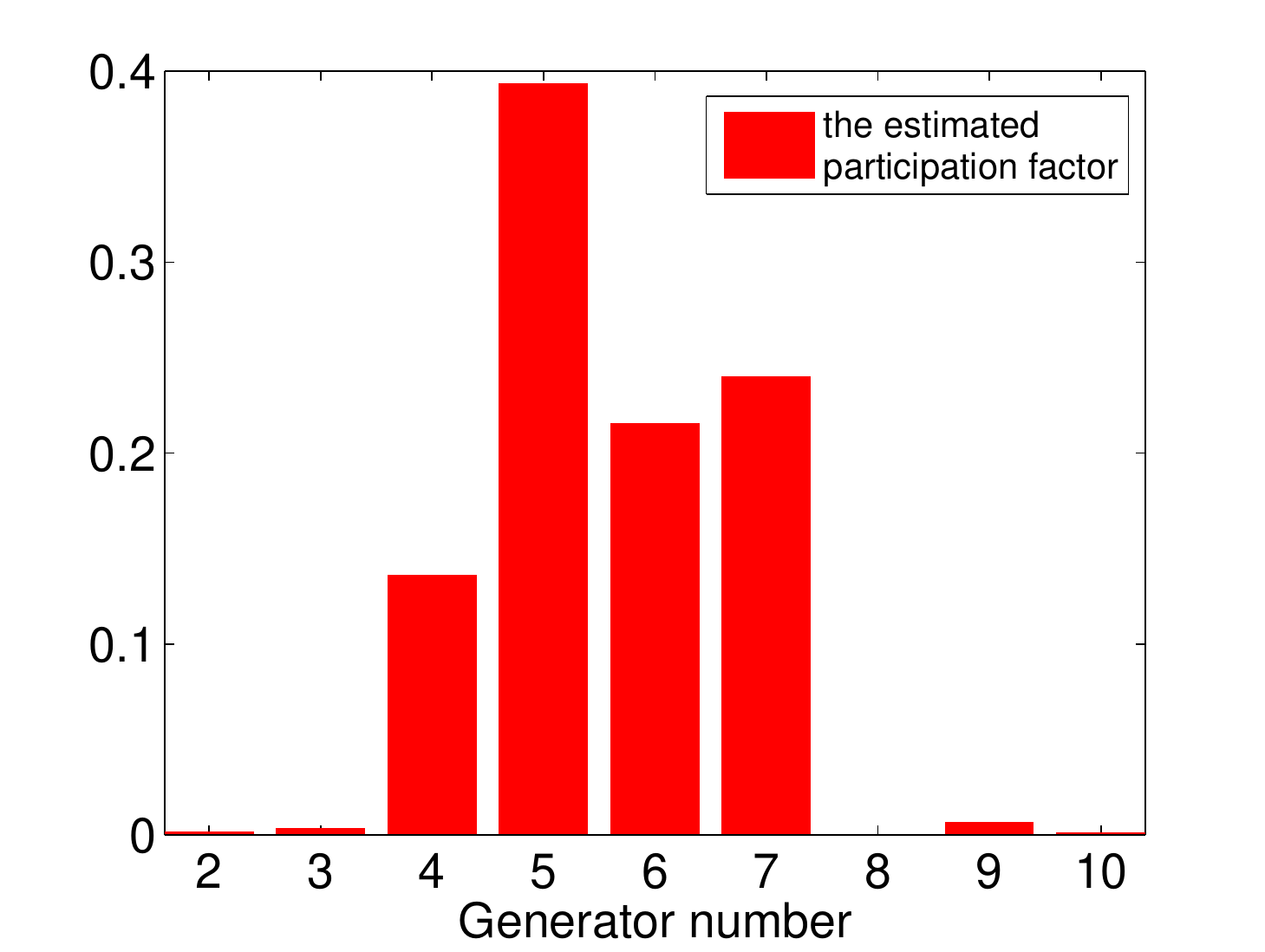}
\caption{Participation factors for Mode 7.}\label{participationfactor-2order}
\end{subfigure}
\caption{Study I: The estimated mode shapes and participation factors for Mode 7.}
\label{Mode shapes and participation factors for Mode 7}
\end{figure}

The estimated mode shapes and participation factors for Mode 7 are presented in Fig. 2. It is clear from Fig. \ref{mode shape mode 7} that Mode 7 is an interarea oscillation mode, in which Generator 4 (G4) and 5 (G5) are swinging against Generator 6 (G6) and 7 (G7). 
The influence of the rest of the generators in Mode 7 is negligible as their participation factors are close to zero.

It is worth noting that the total CPU time needed for the calculation of $A$ 
is 9.642 ms using a computer of 2.50GHz and 8.00GB memory, indicating that the real-time estimation of the system state matrix of the reduced network model is feasible in practical applications.

By the proposed WAMS-based damping control algorithm, the most significant participants in Mode 7, Generator 5-7, 
are chosen to conduct the control. To illustrate how the number of controlled generators influences the damping effect, we perform the following numerical experiments. In the $1^{st}$ experiment, the damping control signal is adopted only at the generator with the largest participation factor---G5. In the $2^{nd}$ experiment, both G5 and G7 receive the damping control signals. In the $3^{rd}$ experiment, we include G6 together with G5 and G7 to apply the control signals. In the $4^{th}$ experiment, all generators participate in the damping feedback loop. The damping coefficient is set to be $\sigma_7 = 2$ in all experiments.
The comparison between the open-loop eigenvalues and the closed-loop eigenvalues is illustrated in Fig. 3. It can be observed that the selected interarea mode gains more damping as the number of connected stations increases. Also, it seems that the exclusion of the generators with negligible participation factors does not have a notable impact on the effectiveness and efficiency of the damping control scheme. 
Moreover, Table \ref{damping_closed} shows that the threshold 10\% is met in the last three experiments, indicating that the proposed technique requires only two generators (G5 and G7) to achieve a desirable damping performance, although more controlled generators will provide an even enhanced damping effect.  
Note that the rest of the modes are not affected by the method.
\begin{figure}[!ht]
\vspace{0pt}
\centering
\begin{subfigure}[t]{0.5\linewidth}
\includegraphics[width=1.7in ,keepaspectratio=true,angle=0]{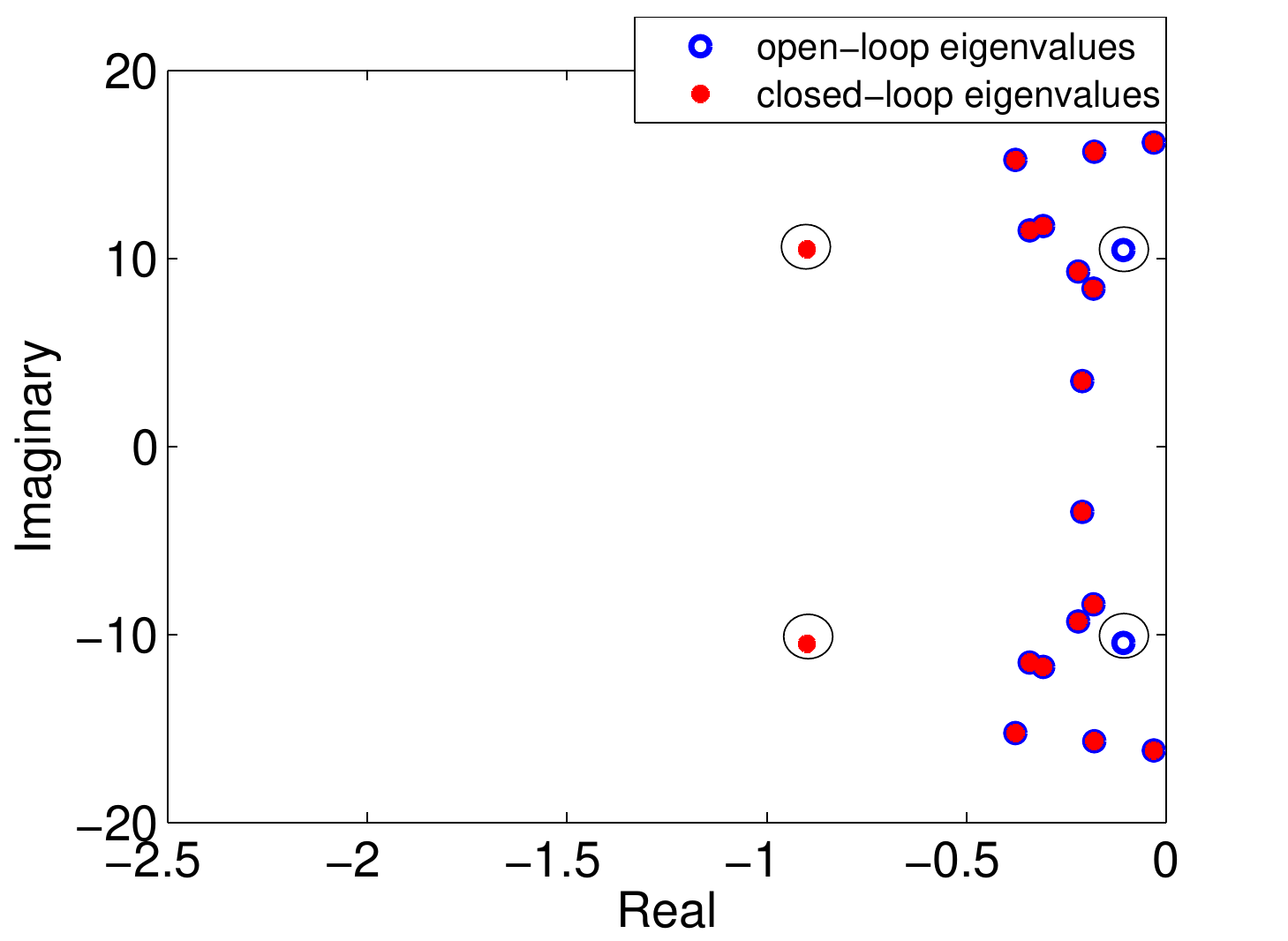}
\caption{Control at G5}
\end{subfigure}%
\begin{subfigure}[t]{0.5\linewidth}
\includegraphics[width=1.7in ,keepaspectratio=true,angle=0]{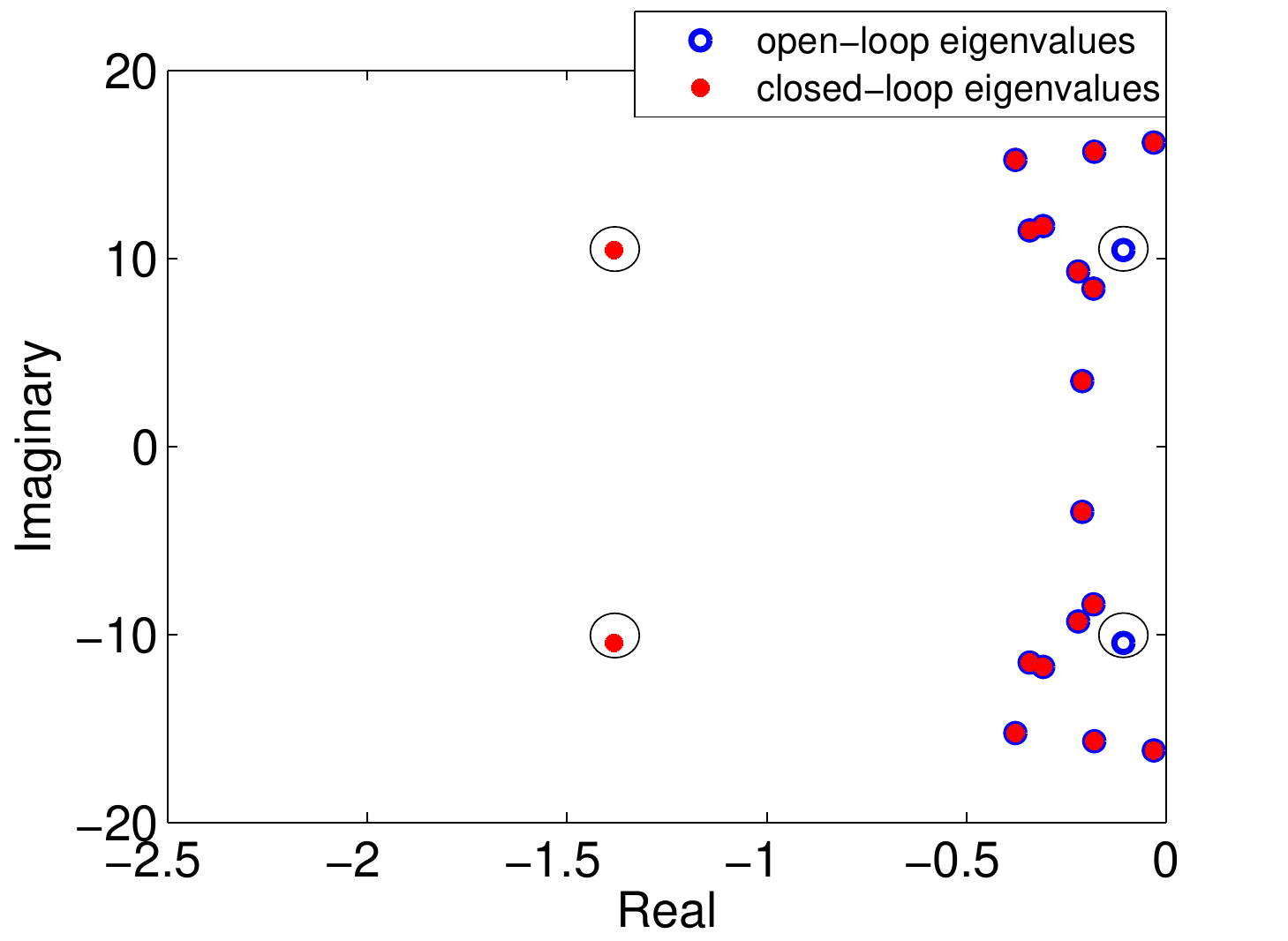}
\caption{Control at G5 and G7}
\end{subfigure}
\begin{subfigure}[t]{0.5\linewidth}
\includegraphics[width=1.7in ,keepaspectratio=true,angle=0]{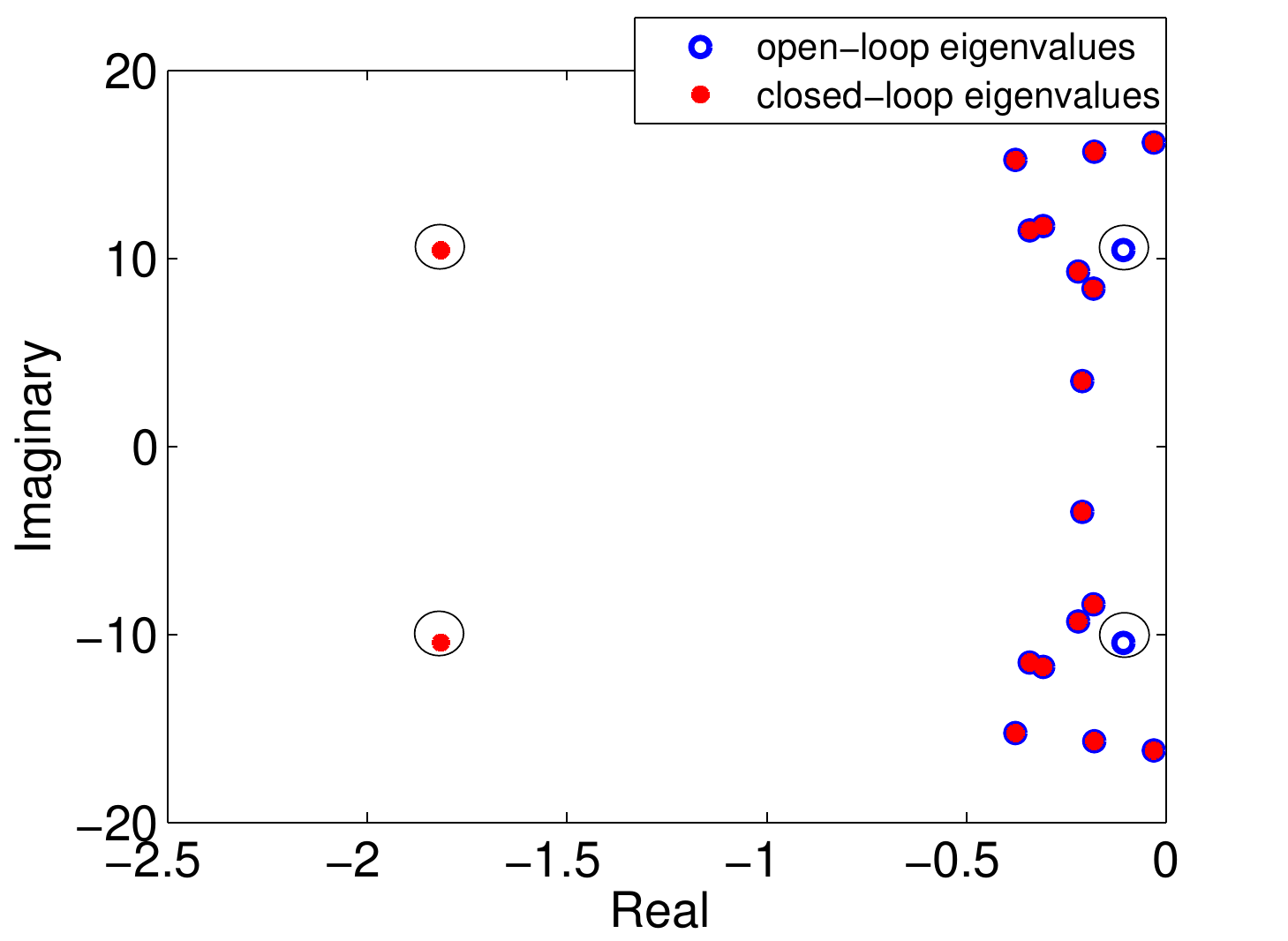}
\caption{Control at G5,G7 and G6}
\end{subfigure}%
\begin{subfigure}[t]{0.5\linewidth}
\includegraphics[width=1.7in ,keepaspectratio=true,angle=0]{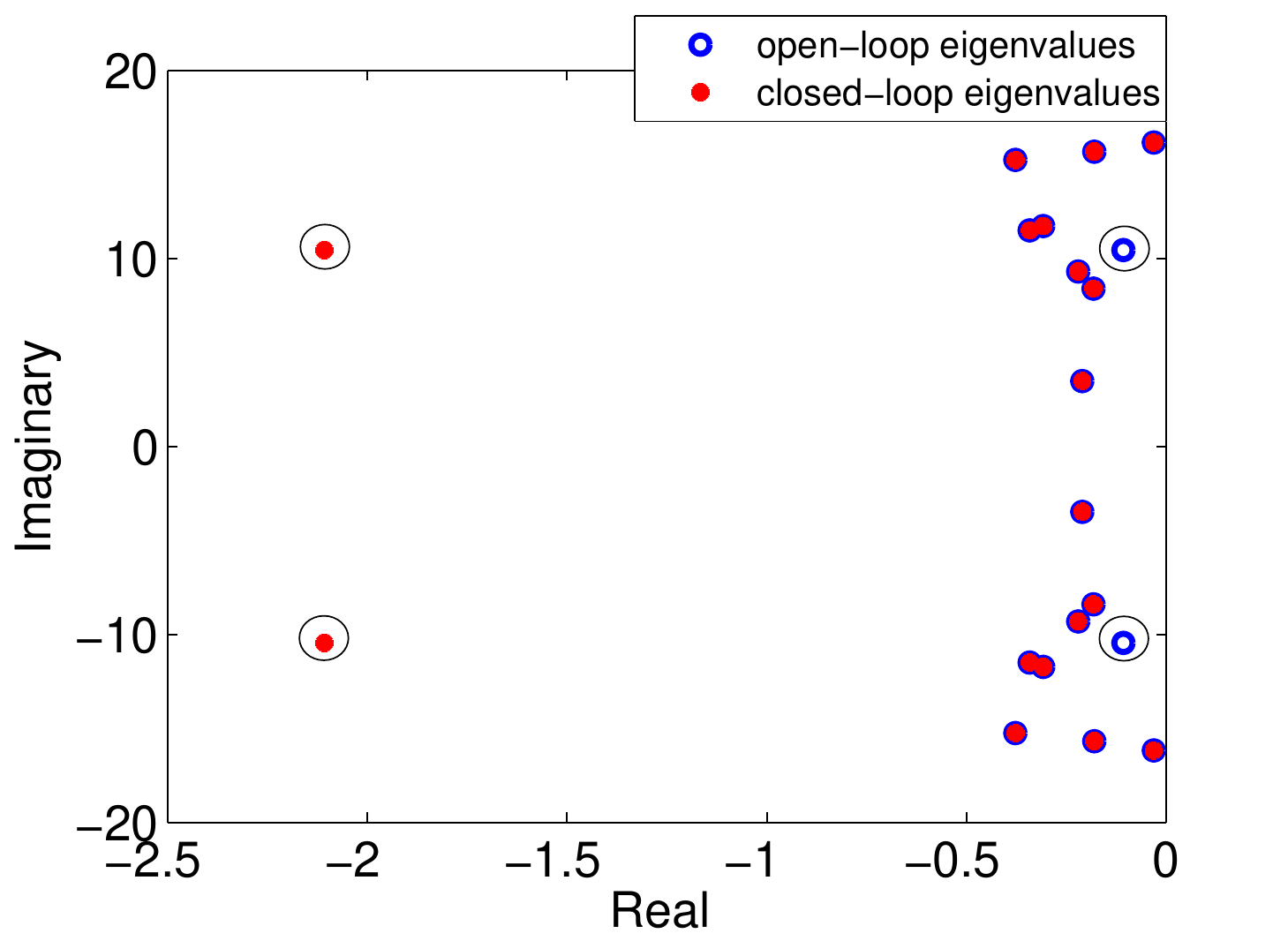}
\caption{Control at all Generators}
\end{subfigure}
\caption{Study I: A comparison between the open-loop and the closed-loop eigenvalues.}
\label{damping_eigenvalues}
\vspace{-6pt}
\end{figure}



\begin{table}[!ht]
\centering
  \caption{Study I: Closed-loop damping ratio for Mode 7.}\label{damping_closed}
  \begin{tabular}{|c|c|}
\hhline{|=|=|}
Generators&Closed-loop \\
  &damping ratio (\%) \\
  \hline
  G5&8.55 \\
  G5, G7&13.11\\
  G5, G7, G6&17.15\\
  All Generators&19.77\\
\hhline{|=|=|}
  \end{tabular}

\vspace{-10pt}
\end{table}


\subsection{Study II: Detailed Generator Model with PMU Measurement Noise}



In this study, all the $10$ generators in the IEEE 39-bus system are modelled by the fourth-order models, which are controlled by field exciters and PSSs. Besides this, a Gaussian-distributed measurement noise with standard deviation of $10^{-3}$ for angles and $10^{-6}$ for rotor speeds is added to the emulated PMU measurements according to the IEEE standard \cite{Amen14}. 

The eigenvalues of the system state matrix $A$ are accurately estimated with an error lower than 2\% for mode frequencies and less than 8\% for damping ratios.
Particularly, Mode 6 that is described by the estimated values $f_6=1.800$ Hz (0.67\% estimation error) and $\zeta_6=2.66\% <10\%$ (3.42\% estimation error) is obviously underdamped.
Fig. 4 presents the estimated mode shapes and participation factors for Mode 6 that is apparently an interarea mode. Indeed, Generator 10 (G10) and 8 (G8) oscillate against Generator 2 (G2) and 9 (G9). These generators take the most responsibility for the excitation of Mode 6.  



\begin{figure}[!ht]
\vspace{0pt}
\centering
\begin{subfigure}[t]{0.5\linewidth}
\includegraphics[width=1.7in ,keepaspectratio=true,angle=0]{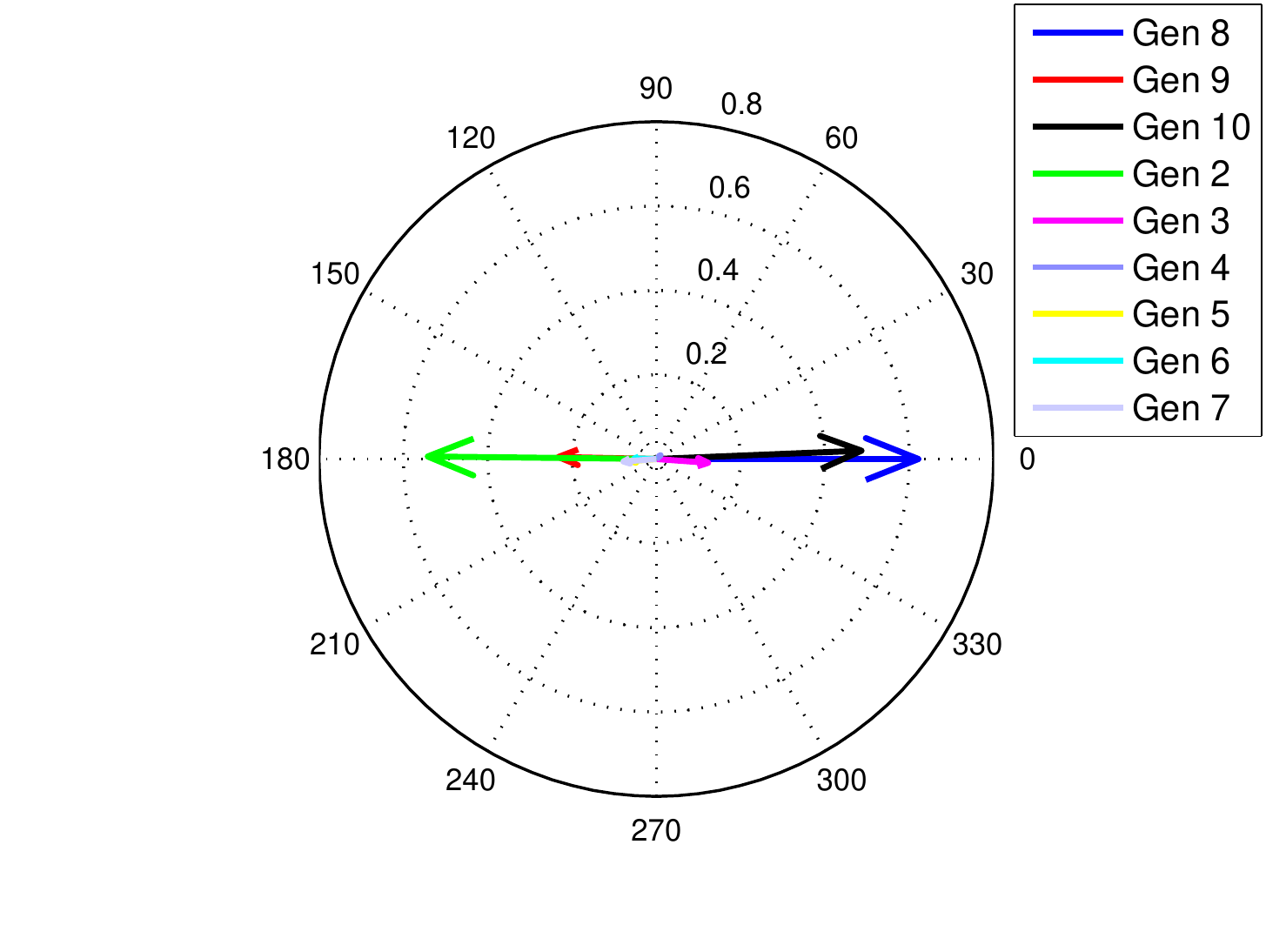}
\caption{Mode shapes for Mode 6.}\label{trajectory}
\end{subfigure}%
\begin{subfigure}[t]{0.5\linewidth}
\includegraphics[width=1.7in ,keepaspectratio=true,angle=0]{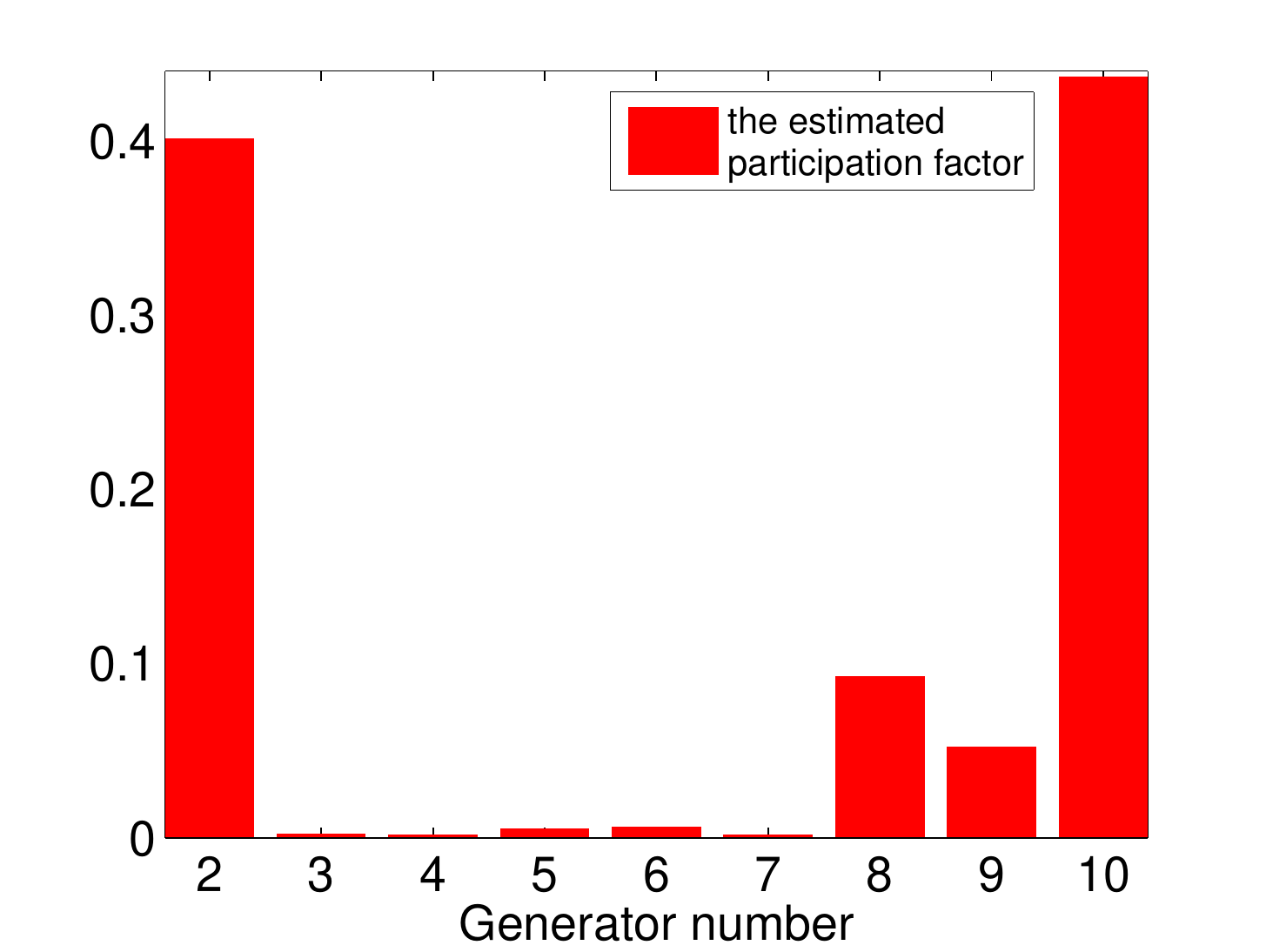}
\caption{Participation factors for Mode 6.}\label{participationfactor-highorder}
\end{subfigure}
\caption{Study II: The estimated mode shapes and participation factors for Mode 6.}
\label{Mode shapes and participation factors for Mode 6}
\end{figure}

It should be noted that in this case the total CPU time needed for the calculation of $A$ 
is 9.731 ms.

The developed damping control technique is implemented by G10, G2 and G8 , the most important participants in Mode 6, utilizing four different experiments. The experiments with an increasing number of controlled generators are designed based on the participation factor ranking while the damping factor is selected to be $\sigma_6 = 2$. The relationship between the open-loop and the closed-loop eigenvalues is shown in Fig. 5. It can be seen that the damping effect increases as the number of the generators participating in the central control grows, which is also corroborated by the damping ratios presented at Table \ref{damping_closed_high}.
Furthermore, the 10\% damping ratio requirement is satisfied by all experiments, implying that the proposed technique can achieve an effective damping impact with only one generator (G10) under control. 

\begin{figure}[!ht]
\centering
\begin{subfigure}[t]{0.5\linewidth}
\includegraphics[width=1.7in ,keepaspectratio=true,angle=0]{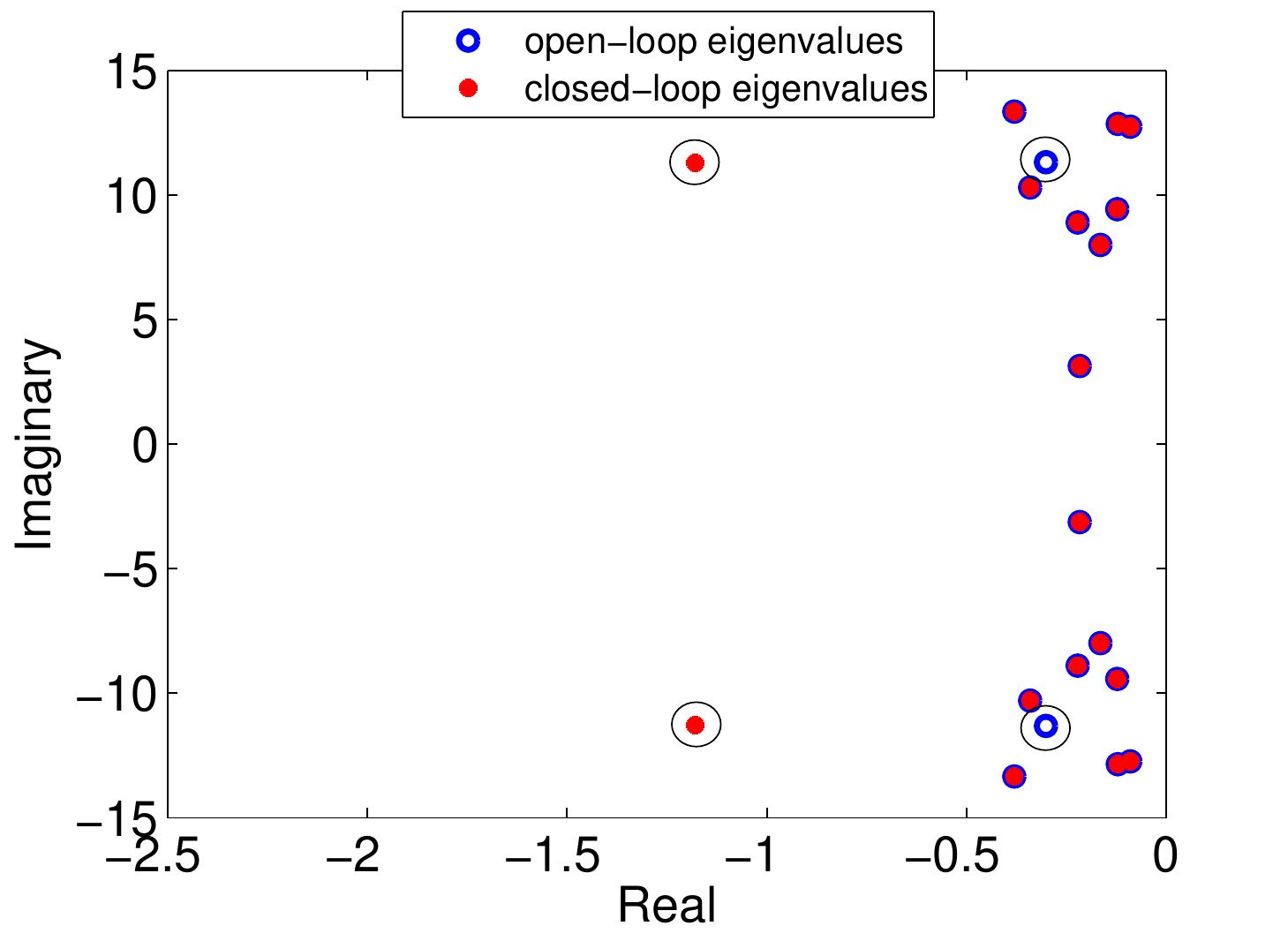}
\caption{Control at G10}
\end{subfigure}%
\begin{subfigure}[t]{0.5\linewidth}
\includegraphics[width=1.7in ,keepaspectratio=true,angle=0]{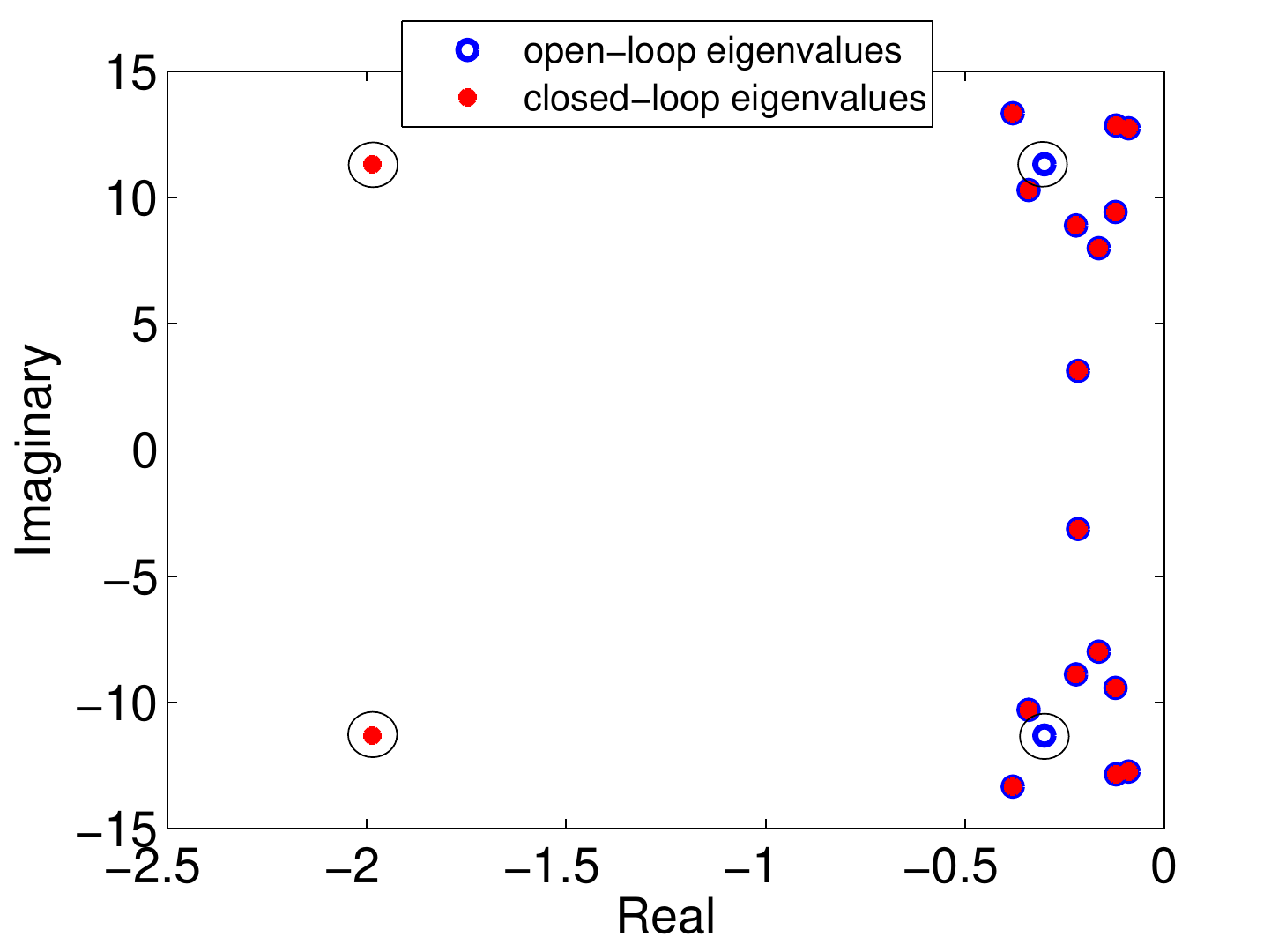}
\caption{Control at G10 and G2}
\end{subfigure}
\begin{subfigure}[t]{0.5\linewidth}
\includegraphics[width=1.7in ,keepaspectratio=true,angle=0]{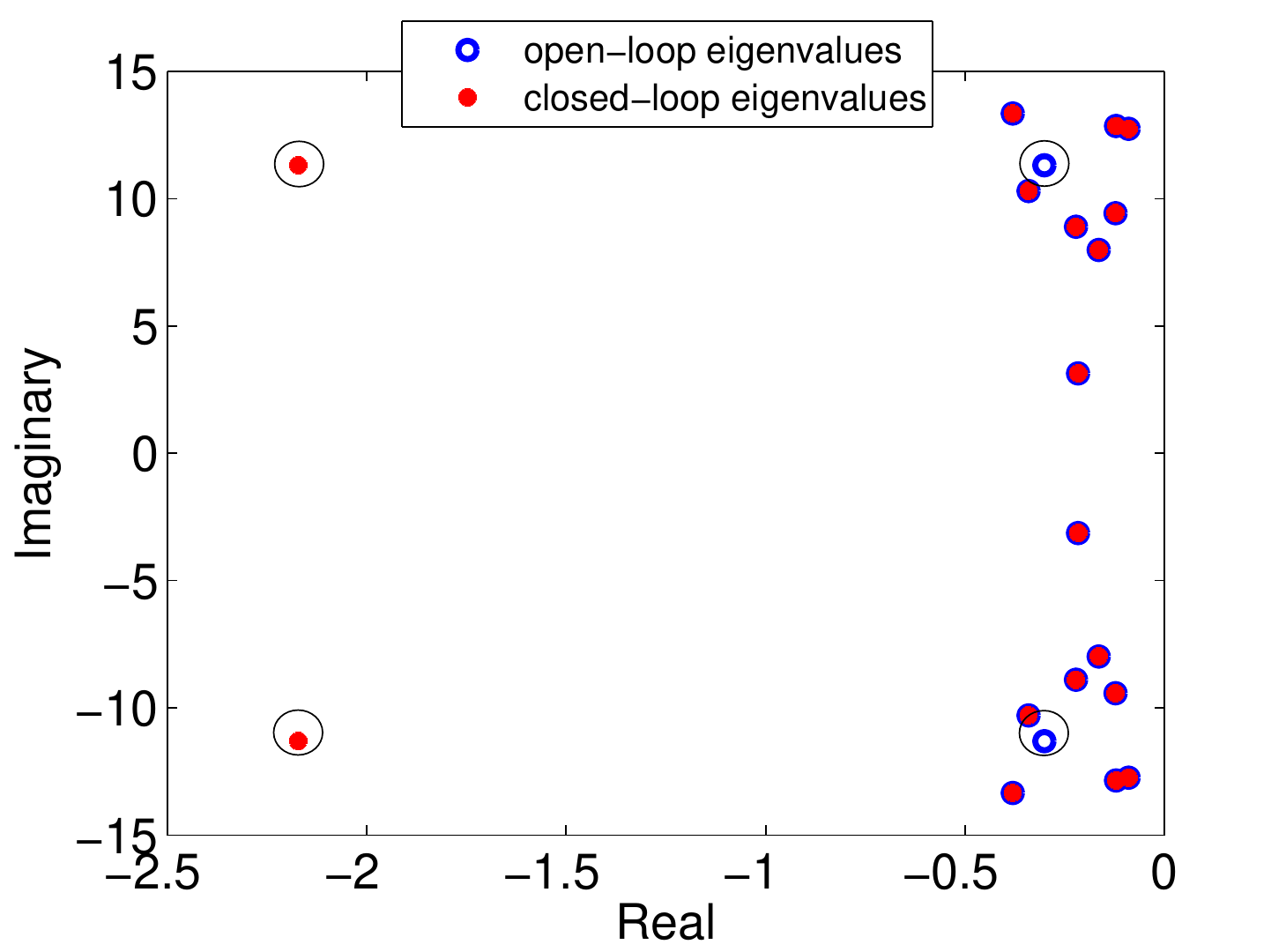}
\caption{Control at G10,G2 and G8}
\end{subfigure}%
\begin{subfigure}[t]{0.5\linewidth}
\includegraphics[width=1.7in ,keepaspectratio=true,angle=0]{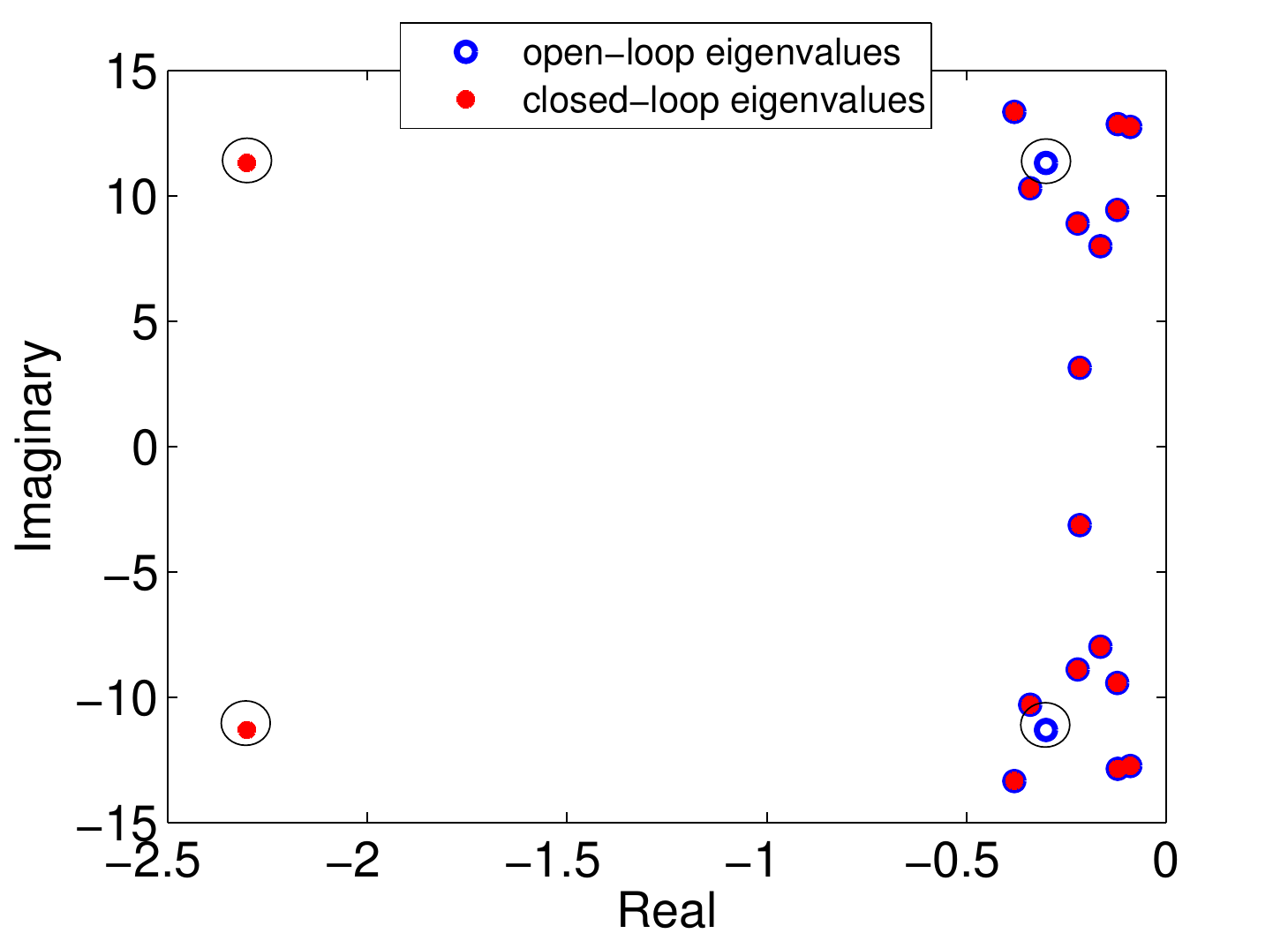}
\caption{Control at all Generators}
\end{subfigure}
\caption{Study II: Comparison between the open-loop and the closed-loop eigenvalues.}
\label{damping_eigenvalues}
\vspace{-6pt}
\end{figure}

\begin{table}[!ht]
\centering
  \caption{Study II: Closed-loop damping ratio for Mode 6.}\label{damping_closed_high}
  \begin{tabular}{|c|c|}
\hhline{|=|=|}

  Generators&Closed-loop \\
  &damping ratio (\%)\\
  \hline
  G10&10.39 \\
  G10, G2&17.29\\
  G10, G2, G8&18.86\\
  All Generators&19.93\\
\hhline{|=|=|}
  \end{tabular}
\vspace{0pt}
\end{table}

\begin{figure}[!ht]
\centering
\includegraphics[width=2in ,keepaspectratio=true,angle=0]{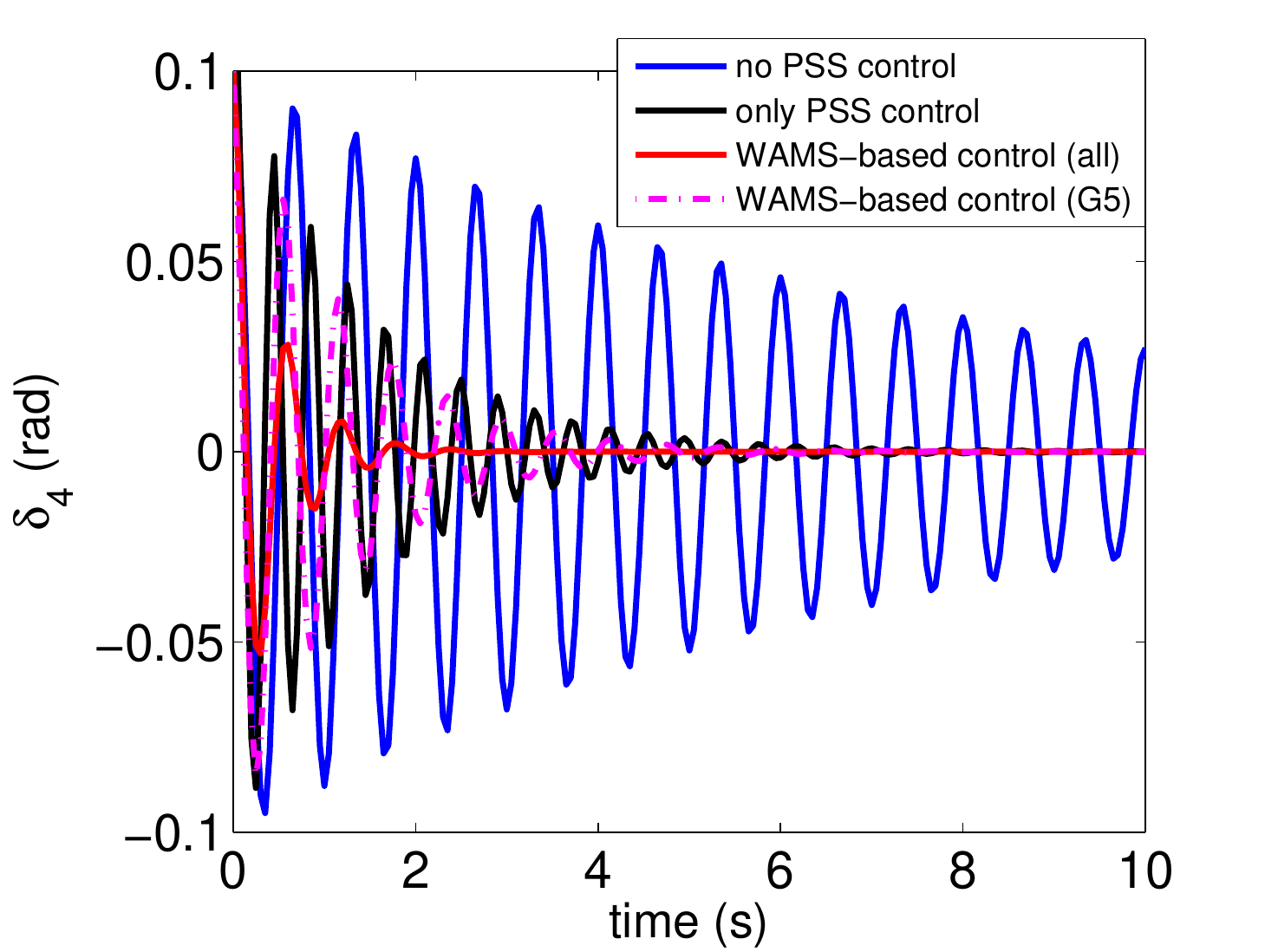}
\caption{Study II: Time-domain response of $\delta_4$ to the excitation of Mode 5.}\label{response2}
\vspace{-14pt}
\end{figure}

The last but not the least, the effectiveness of the proposed approach is demonstrated by comparing it with the conventional PSS technique. Particularly, the interarea Mode 7 ($f_7=1.662$ Hz and  $\zeta_7=1.03\%$) was excited and the time-domain response of $\delta_4$ (Fig. \mbox{\ref{response2}}) was simulated for the following cases:
\begin{itemize}
\item Case A: No PSS control;
\item Case B: PSS control at all generators;
\item Case C: WAMS-based control at all generators;
\item Case D: WAMS-based control at G5 (biggest participation factor);

\end{itemize}

It can be seen that the proposed method (Case C and Case D) achieves an improved damping performance compared to Case A and Case B. 
Even when the control is conducted at only one generator (G5), the damping performance of the proposed wide-area damping method is better than the PSS local control. As known, the conventional PSS is only effective in a typically narrow frequency range. Although multi-band PSS\mbox{\cite{Lajoie98}} may enhance the performance, a complicated tuning process is required and may affect the rest of the modes. In addition, these approaches may not work well if the assumed network model is subject to constant changes. 
In contrast, the proposed wide-area damping control method can effectively damp the target mode by any selected damping coefficient using a small number of generators while maintaining the other modes unaffected. More importantly, the network model and parameter values are not assumed to be known.

\vspace{0pt}
\section{CONCLUSIONS AND PERSPECTIVES}\label{4}
This paper proposes a wide-area damping control method using PMU data to damp the undesirable interarea oscillations in the modern power grid. The proposed method does not depend on the network model and can be integrated into online dynamic security assessment (DSA) for continuous monitoring and controlling the interarea oscillations. It has been shown analytically and numerically that the targeted mode can be adequately damped using a small number of synchronous machines. 
In the future, our efforts will focus on simultaneously damping multiple interarea modes in larger power systems exploiting the estimated participation factors.

\vspace{0pt}

\appendix
\label{appendix}
\section{Determine the Covariance Matrix}
The stationary covariance matrix is given by:
\begin{eqnarray}
\label{eq:stat_cov}
C_{\bm{x}\bm{x}}&=&
\begin{bmatrix}
C_{\bm{\delta\delta}}&C_{\bm{\delta\omega}}\\
C_{\bm{\omega\delta}}&C_{\bm{\omega\omega}}\\
\end{bmatrix}\nonumber
\end{eqnarray}
%
where, for instance, $C_{\delta_{i}\delta_{j}} = \mathbf{E}[(\delta_{i}-\mu_{i})(\delta_{j}-\mu_{j})]$, and $\mu_i$ is the mean of $\delta_i$. In practice, $C_{\bm{\delta\delta}}$ is usually unknown due to insufficient data. Thus, $C_{\bm{\delta\delta}}$ is estimated by the sample covariance matrix $Q_{\bm{\delta\delta}}$, the $(i,j)^{th}$ element of which is computed as \cite{Gardiner09}:
$Q_{\delta_{i}\delta_{j}} = \frac{1}{N - 1}\sum\limits_{k=1}^N (\delta_{ki}-\bar{\delta_{i}})(\delta_{kj}-\bar{\delta_{j}})
$,
%
where $\bar{\delta}_{i}$ symbolizes the sample mean of $\delta_i$, and $N$ is the sample size. Similarly, $Q_{\bm{\omega}\bm{\omega}}$ and $Q_{\bm{\omega}\bm{\delta}}$ are used to estimate $C_{\bm{\omega}\bm{\omega}}$ and $C_{\bm{\omega}\bm{\delta}}$ respectively. A window size of $450$s is used in the examples of this paper, which shows good accuracy.

It should be noted that the proposed technique of estimating $C_{\bm{x}\bm{x}}$ is fast and efficient as shown in the simulation study. 
Moreover, $C_{\bm{x}\bm{x}}$ can be estimated recursively using
a fast iterative approach, which will further reduce the computational effort.

\vspace{-10pt}
%
%
%
%
%
%
%

\end{document}